\documentclass[showkeys,useAMS,nofootinbib,pra,twocolumn,superscriptaddress]{revtex4-1}
\usepackage[T1]{fontenc}
\usepackage[latin9]{inputenc}
\usepackage{graphics,graphicx}
\usepackage{multirow}
\usepackage{booktabs}
\usepackage[table,xcdraw]{xcolor}
\usepackage[breaklinks=true]{hyperref}
\usepackage{breakcites}
\usepackage{mathtools}
\usepackage{amsmath}
\usepackage{amsthm}
\usepackage{amssymb}
\usepackage{xcolor}
\usepackage{mathrsfs}
\usepackage{upgreek}
\usepackage{soul}
\usepackage[normalem]{ulem}

\makeatletter
\theoremstyle{plain}

\usepackage{dsfont}
\newcommand{\id}{\mathds{1}}

\newcommand{\tr}{\text{Tr}}

\newcommand{\eq}[1]{\begin{equation}
    \begin{aligned}#1\end{aligned}
\end{equation}}
\newcommand{\ket}[1]{\left|#1\right\rangle}
\newcommand{\bra}[1]{\left\langle#1\right|}
\newcommand{\braket}[2]{\left\langle#1\bigg|#2\right\rangle}
\newcommand{\expct}[1]{\left\langle#1\right\rangle}

\newcommand{\iu}{\text{i}}
\newcommand{\eu}{\text{e}}

\newcommand{\had}{\hat{a}^\dagger}

\newcommand{\add}[1]{#1}
\newcommand{\rem}[1]{}

\makeatother

\usepackage[english]{babel}

\providecommand{\theoremname}{Theorem}

\makeatletter
\newcommand{\printfnsymbol}[1]{%
  \textsuperscript{\@fnsymbol{#1}}%
}
\makeatother

\begin{document}

\title{
Multiphase estimation without a reference mode
}
\author{Aaron Z. Goldberg}
\email{goldberg@physics.utoronto.ca}
\affiliation{Department of Physics and Centre for Quantum Information \& Quantum Control, University of Toronto, Toronto, Ontario, Canada M5S 1A7}
\author{Ilaria Gianani}
\affiliation{Dipartimento di Fisica, Sapienza Universit\`a di Roma, Roma, Italy, I-00185}
\affiliation{Dipartimento di Scienze, Universit\`a degli Studi Roma Tre, Via della Vasca Navale 84, 00146 Rome, Italy}
\author{Marco Barbieri}
\affiliation{Dipartimento di Scienze, Universit\`a degli Studi Roma Tre, Via della Vasca Navale 84, 00146 Rome, Italy}
\author{Fabio Sciarrino}
\affiliation{Dipartimento di Fisica, Sapienza Universit\`a di Roma, Roma, Italy, I-00185}
\author{Aephraim M. Steinberg}
\affiliation{Department of Physics and Centre for Quantum Information \& Quantum Control, University of Toronto, Toronto, Ontario, Canada M5S 1A7}
\affiliation{CIFAR, 661 University Ave., Toronto, Ontario M5G 1M1, Canada}
\author{Nicol\`o Spagnolo}
\affiliation{Dipartimento di Fisica, Sapienza Universit\`a di Roma, Roma, Italy, I-00185}

\begin{abstract}
    Multiphase estimation is a paradigmatic example of a multiparameter problem. When measuring multiple phases embedded in interferometric networks, specially-tailored input quantum states achieve enhanced sensitivities compared with both single-parameter and classical estimation schemes. Significant attention has been devoted to defining the optimal strategies for the scenario in which all of the phases are evaluated with respect to a common reference mode, in terms of optimal probe states and optimal measurement operators. As well, the strategies assume unlimited external resources, which is experimentally unrealistic.
    Here, we optimize a generalized scenario that treats all of the phases  on an equal footing and takes into account the resources provided by external references. We show that the absence of an external reference mode reduces the number of simultaneously \rem{estimatable}\add{estimable} parameters, owing to the immeasurability of global phases, and that the symmetries of the parameters being estimated dictate the symmetries of the optimal probe states. Finally, we provide insight for constructing optimal measurements in this generalized scenario. The experimental viability of this work underlies its immediate practical importance beyond fundamental physics.
\end{abstract}

\maketitle

\section{Introduction}
 With its potential to revolutionize fields such as imaging and sensing, quantum metrology is one of the most promising near-term quantum technologies. Photonics implementations are prominent, with many problems cast as the measurement of a single optical phase, whose applications range from the measurement of biological tissues~\cite{Nolte2012} to the detection of gravitational waves~\cite{LIGO2011,Abbottetal2016}. For such tasks, the advantage of using quantum light is a long-established result~\cite{Caves1981,BondurantShapiro1984,Yurkeetal1986,Grangieretal1987,Xiaoetal1987} and a long-sought technological goal \cite{PhysRevLett.107.113603,Nagata726,Higgins:2007xy,Kacprowicz:2010jk, Slussarenko:2017qv,doi:10.1063/1.4724105}.
 
 However, this focus on the single-parameter case is neither necessary nor advisable. Recent suggestions advise adopting a multiple parameter approach~\cite{Rolkeetal2005,Suzuki2020}, thus making quantum-enhanced multiparameter estimation \cite{Humphreysetal2013,Szczykulskaetal2016,BaumgratzDatta2016,Ragyetal2016,Rehaceketal2017,Chrostowskietal2017,GoldbergJames2018,Liuetal2019,Polino:s,SidhuKok2020,Rubioetal2020,Albarellietal2020} an important component of the next quantum revolution. 

The paradigmatic
multiparameter estimation problem is the
estimation of multiple relative phases in an interferometer.
This proof-of-concept scenario, in which a simultaneous estimation strategy can outperform sequential quantum-enhanced estimation strategies, has generally been approached assuming the presence of a preferred reference mode and an equal interest in the remaining $d$ modes \cite{Humphreysetal2013}, although different choices have also been considered~\cite{Knottetal2016}. 

In this work, we present a comprehensive study of the implications of the presence or absence of a phase reference for multiple phase estimation, extending the results of Ref\add{s}.~\cite{JarzynaDemkowiczDobrzanski2012}\add{\cite{Ataman2020}}. Our treatment of this estimation problem makes use of the Quantum Fisher Information (QFI). This is a powerful tool encapsulating the ultimate lower bound on the precision that can be achieved for estimating a specific parameter using a given state \cite{Helstrom1967,Holevo1973}. In multiparameter problems, the corresponding QFI becomes a matrix,
whose inverse bounds the matrix of covariances between all of the estimated parameters. Scalar versions of the bound can be inferred, limiting the precision of estimating all parameters simultaneously.

The QFI, however, does not take into account any experimental restrictions on what measurements can be \textit{feasibly} achieved, leaving the theoretical treatment of estimation somewhat disconnected from practical considerations. By explicitly incorporating the availability of a phase reference, we are able to transparently include this experimental resource into the QFI framework \add{without recourse to infinitely intense reference beams}. 

This article is organised as follows: in Section II, we summarise our main results; in Section III, we introduce the multiphase paradigm; in Section IV, we detail the treatment of multiple phase estimation with classical light; in Section V, we extend our studies to quantum states of light; and in Section VI, we address experimental implementations of the required measurements.

\section{Summary of the results}

\begin{figure}[htbp]
    \centering
    \includegraphics[width = \columnwidth]{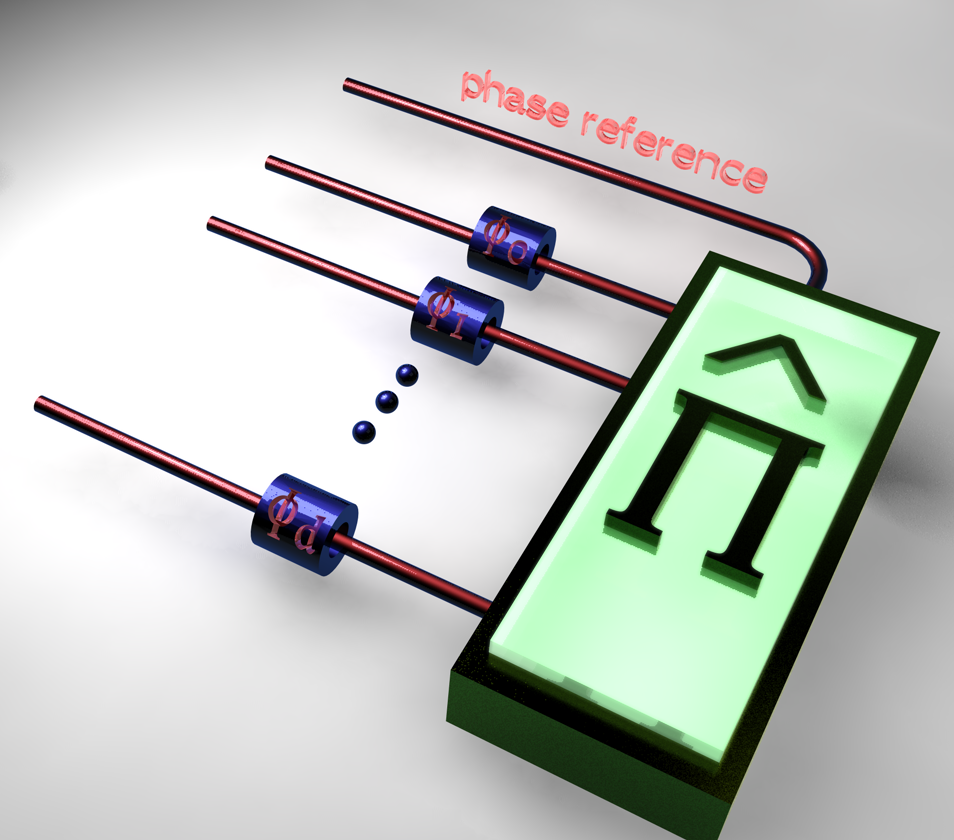}
    \caption{Multiple phase estimation: general concept. A set of optical phases $\phi_0,\cdots,\phi_d$ is estimated, based on a measurement strategy $\Pi$. This can either make use of an external phase reference, or use one or more of the modes as a reference.}
    \label{fig:setup}
\end{figure}

The imprinting of a phase shift $\phi$ on an optical mode is described mathematically by the action of the operator $\eu^{\iu\phi \hat{n}}$, where $\hat{n}$ is the photon number operator acting on that mode. Operating on a Fock state $\ket{n}$, the phase shift operator yields $\eu^{\iu n\phi}\ket{n}$, where the parameter $\phi$ appears only as an unobservable global phase. In contrast, coherent states $\ket{\alpha}\propto\sum_{n=0}^\infty \tfrac{\alpha^n}{\sqrt{n!}}\ket{n}$, the most classical states of light, transform as $\ket{\alpha}\to\ket{\eu^{\iu\phi}\alpha}$, the phase $\phi$ now being encoded as a relative phase between the amplitudes of the component number states, which is, in principle, measurable.

\begin{figure*}[ptb!]
    \centering
    \includegraphics[width = \textwidth]{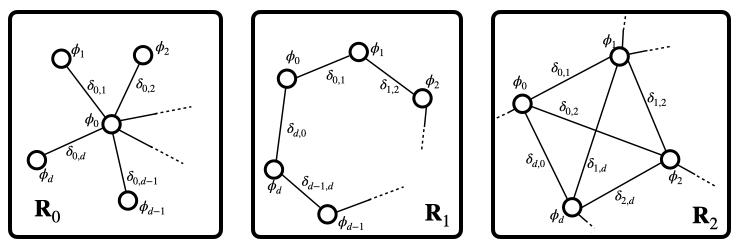}
    \caption{Schematics of the different phase estimation strategies. Circles stand for the phases in each mode, and the connecting lines for the parameters to be estimated. In the first panel, mode 0 is selected as a privileged phase reference, corresponding to estimating the relative phases $\delta_{i,0}=\phi_i-\phi_{0}$. In the second panel, the choice is to refer each phase to the previous one (in cyclical fashion): $\delta_{i,j}=\phi_i-\phi_{j}$. Finally, in the third panel, all relative phases are considered.}
    \label{fig:schemes}
\end{figure*}

{ However, interfering different energy eigenstates in order to measure their relative phase is only possible with the aid of an ancillary system with uncertain energy: here, another beam $\ket{\beta}$, which can be used as a phase reference.} The necessity of a reference beam is normally glossed over in phase estimation protocols. {As a first result, we quantify how the available information is decreased in the absence of such a phase reference: the rank of the QFI matrix (QFIM) is decreased by one, making it impossible to simultaneously estimate all $d+1$ phases in Fig. 1.} The rank of the QFIM immediately dictates the number of  \add{independent} parameters that can be estimated \add{using a particular probe state}, and the scaling of the QFIM with various experimental parameters informs the metrological usefulness of \rem{a}\add{the} given quantum state. We use this to explicitly show why global phases cannot be estimated in multiphase estimation protocols. 

The irrelevance of global phases implies that no phase is more equal than others. 
This amounts, in practice, to establishing a phase reference, based on the available modes. The parameters to be estimated are thus not the original phases, but  {some set of linear combinations thereof, determined according to a cost matrix $\mathbf{R}_i$.} \add{Our results show how to properly account for any chosen reference beam, avoiding accidental assumptions of access to ancillary beams with infinite energy. This is particularly relevant, for instance, for integrated optical sensors, which often incur severe power limitations. }

We establish the classical limits pertaining to the different scenarios illustrated in Fig.~\ref{fig:schemes}: the standard approach of selecting one of the modes as a reference and estimating the relative phases of the other $d$ modes to that one; the estimation of the $d+1$ relative phases between ``neighboring'' modes; and the estimation of all $d(d+1)/2$ possible relative phases. Some of these have
more parameters than the original problem, but each represents a meaningful task\add{, and none have more than $d$ \textit{independent} parameters}. The optimal partitioning of the total available energy $E$ among the different modes depends on the cost matrix: in particular, we find that for the scenarios $\mathbf{R}_1$ and $\mathbf{R}_2$ a symmetric subdivision begets optimal performance. \add{The optimal measurement strategy simultaneously estimates all of the \textit{interdependent} parameters by inferring them from a projection-valued measure that it set by the optimal probe state.}

 We then discuss how generalized $N00N$ states spread over multiple modes lead to a scaling enhancement of the total variance with respect to the classical limits, generalizing the results of Refs. \cite{Humphreysetal2013,Pezzeetal2017}. We find that the weights of the different components of the quantum state closely follow the prescriptions for classical light.

In particular, an egalitarian estimation scheme optimizes the sensitivity of measuring all phases relative to each other.
In such a scenario, the optimal states have the same form for both classical and nonclassical states, with equal energy in each of the modes of the interferometer. Simultaneous estimation schemes significantly outperform sequential schemes for symmetrized measurement scenarios.

\section{The multiphase paradigm of quantum estimation}

The goal of parameter estimation is to measure a set of parameters $\pmb{\phi}=\left(\phi_0,\cdots,\phi_d\right)$ describing a sample with as much precision as possible. For this purpose, a probe is prepared in  a suitable quantum state $\vert \psi_0\rangle$, which is then transformed by a unitary operation $\hat U(\pmb{\phi})$, representing the action of the sample. {Finally, appropriate measurements are carried out on the output state $\vert \psi(\pmb{\phi})\rangle=\hat U(\pmb{\phi})\vert \psi_{0}\rangle$, so that the values of the parameters can be inferred from the outcome statistics.} 

For the specific case of $d+1$ optical phases, sketched in Fig.\ref{fig:setup}, the explicit form of the unitary operator is
 \eq{
\hat{U}\left(\pmb{\phi}\right)=\exp\left(\iu\sum_{i=0}^d\phi_i\hat{n}_i\right),
} where $\hat{n}_i$ is the photon number operator pertaining to the mode labeled by $i$.

The precision in such a multiparameter case is captured by the $(d+1)\times(d+1)$ covariance matrix with components $\pmb{C}_{i,j}=\left\langle \phi_i \phi_j\right\rangle-\left\langle \phi_i \right\rangle \left\langle\phi_j\right\rangle$.
This is bounded by the Cram\'er-Rao inequality:
\eq{
\pmb{C}\geq \left[\pmb{H}\left(\psi_0;\pmb{\phi}\right)\right]^{-1},
\label{eq:qCRB matrix}
}
where $\pmb{H}\left(\psi_0;\pmb{\phi}\right)$ is the celebrated quantum Fisher information matrix (QFIM). This is a measure of the amount of information about the set $\pmb{\phi}$ that can be extracted from the probe state $\vert \psi_0 \rangle$. In our specific example, the QFIM has components 
\eq{
\pmb{H}_{i,j}&=4\Re\left[
\braket{\frac{\partial \psi(\pmb{\phi})}{\partial \phi_i}}{\frac{\partial \psi(\pmb{\phi})}{\partial \phi_j}}  \right. \\  &\quad\left.-
\braket{\psi(\pmb{\phi})}{\frac{\partial \psi(\pmb{\phi})}{\partial \phi_i}}
\braket{\frac{\partial \psi(\pmb{\phi})}{\partial \phi_j}}{\psi(\pmb{\phi})}\right]\\
&=4\text{Cov}_{\psi(\phi)}(\hat{n}_i,\hat{n}_j),} where \eq{\text{Cov}_{\psi}(X,Y)&=\tfrac{1}{2}\bra{\psi}{XY+YX}\ket{\psi}-\bra{\psi}X\ket{\psi}\bra{\psi}Y\ket{\psi},} 
while the general expression can be found in~\cite{Paris2009}.

We seek to maximize $\pmb{H}$ for the sake of obtaining the minimal covariance, with all the  
caveats of working with a matrix inequality such as \eqref{eq:qCRB matrix}; foremost, we require a scalar figure of merit~\cite{Albarellietal2019}. 

In this framework, it is in principle possible to simultaneously estimate all $d+1$ parameters $\pmb{\phi}$: for some states $\ket{\psi_0}$ the QFIM $\pmb{H}$ has rank $d+1$ so that the lower bound from \eqref{eq:qCRB matrix} is finite\add{, implying that the states independently depend on all $d+1$ parameters}. This does not amount to measuring absolute phases because the QFIM assumes the unconditional availability of an \add{extra} external phase reference, as in Fig.~\ref{fig:setup}. This may not be the case in actual experiments, thus the QFIM calculated using pure states would lead to a too-generous estimate of the attainable covariance, in line with the considerations of Ref.~\cite{JarzynaDemkowiczDobrzanski2012} for single-phase interferometry. 

In the absence of an external phase reference, the appropriate result can be obtained by considering the QFIM derived under a superselection rule that transforms the state in such a way that it erases any global phase information:
\eq{\ket{\psi}\bra{\psi}\to &\int \frac{d\theta}{2\pi}\eu^{-i\hat{N}\theta}\ket{\psi}\bra{\psi}\eu^{+i\hat{N}\theta}\\
&=\sum_{N=0}^\infty\hat{\id}_N\ket{\psi}\bra{\psi}\hat{\id}_N\equiv\sum_{N=0}^\infty p_N\ket{\psi_N}\bra{\psi_N},\label{eq:superselection rule}}
where $\hat{N}=\sum_N N \hat{\id}_N$ is the total-photon-number operator and $\hat{\id}_N$ is the projector onto the $N$-photon subspace. Calculating the QFIM with this transformed state yields the maximal possible experimental precision in the absence of any additional external resources. Since the \add{spans of the} subspaces and their weights $p_N$ \rem{are not changed by varying} \add{do not depend on} $\pmb{\phi}$, the resulting mixed-state QFIM is the convex sum of the corresponding pure-state QFIMs (\cite{SidhuKok2020}):
\eq{\pmb{H}\left(\sum_{N=0}^\infty p_N\ket{\psi_N}\bra{\psi_N}\right)=\sum_{N=0}^\infty p_N\pmb{H}\left(\ket{\psi_N}\bra{\psi_N}\right).}

We can then demonstrate our first result: this superselected QFIM $\pmb{H}$ has rank at most $d$ \add{and thus \textit{no} state can independently depend on more than $d$ parameters}. For this purpose, we observe that the new QFIM $\pmb{H}$ can be broken into a convex sum: 
\eq{
\pmb{H}_{i,j}=4\sum_N p_N\text{Cov}_{\ket{\psi_N}}(\hat{n}_i,\hat{n}_j).
} In each photon-number subspace, we can rewrite $\hat{n}_0=N-\sum_{i=1}^d\hat{n}_i$, where $N$ is a constant. Using the linear covariance rule $\text{Cov}_{\ket{\psi}}(X,Y+Z)=\text{Cov}_{\ket{\psi}}(X,Y)+\text{Cov}_{\ket{\psi}}(X,Z)$, we now find
\eq{
\pmb{H}_{i,0}=\pmb{H}_{0,i}
&=\sum_{j=1}^d\left[-4\sum_N p_N \text{Cov}_{\ket{\psi_N}}(\hat{n}_i,\hat{n}_j)\right]} and
\eq{
\pmb{H}_{0,0}
&=\sum_{i,j=1}^d4\sum_N p_N\text{Cov}_{\ket{\psi_N}}(\hat{n}_i,\hat{n}_{\rem{i}\add{j}})
.
} 
From this it is apparent that $\sum_{j=0}^d\pmb{H}_{i,j}=0$ for all $i$, from which we immediately conclude that $\pmb{H}$ is singular. 

\add{
A scalar version of the bound \eqref{eq:qCRB matrix} can be found using any positive-definite cost matrix $\mathbf{R}$:
\eq{
\tr\left(\mathbf{R}\pmb{C}\right)\geq \tr \left\{\mathbf{R}\left[\pmb{H}\left(\psi_0;\pmb{\phi}\right)\right]^{-1}\right\}.
\label{eq:scalar qCRB}
} Given that that $\pmb{C}$ and $\pmb{H}$ are symmetric, we can take $\mathbf{R}=\pmb{J}^T\pmb{J}$ to be symmetric without loss of generality for some real matrix $\pmb{J}$. Any measurement that saturates the matrix inequality \eqref{eq:qCRB matrix}, which is not possible in general but is always possible for multiphase estimation, also saturates the scalar inequality \eqref{eq:scalar qCRB} for a particular probe state. Nonetheless, different probe states yield different lower bounds for different cost functions. Thus, one first defines a cost function $\mathbf{R}$ based on physical considerations of the relevant parameters to be estimated, weighing each parameter by its relative significance, which can be done regardless of the basis in which one originally parametrizes the problem; next, one searches for a probe state that minimizes the lower bound of the scalar inequality \eqref{eq:scalar qCRB}; then, finally, one searches for a measurement procedure that will saturate the matrix inequality \eqref{eq:qCRB matrix}. The measurement procedure yields the covariance matrix $\pmb{C}$, with which one can infer any cost function, and this procedure saturates the scalar inequality for \textit{any} cost matrix; still, the probe state is only guaranteed to be optimal for a \textit{particular} cost matrix.
}

\section{Optimal estimation with classical states}

The capabilities of multiphase estimation with classical light are assessed by inspecting the state $\bigotimes_{i=0}^d\ket{\eu^{\iu\phi_i}\alpha_i}$. In the presence of a phase reference, the associated QFIM has components
\eq{\pmb{H}_{i,j}=4\left|\alpha_i\right|^2\pmb{\updelta}_{i,j},\label{eq:QFIM coherent with reference}} 
where  $\pmb{\updelta}_{i,j}$ is the Kronecker delta. 
The diagonal form of this QFIM derives from  the fact that each phase shift is accumulated independently of the others. This implies that, in principle, each phase $\phi_i$ can be estimated at its individual ultimate limit, regardless of the presence of the others.
The corresponding variances will be proportional to the inverses of the energies in each mode $i$.

When a phase reference is unavailable, the form \eqref{eq:QFIM coherent with reference} is not valid, and needs to be replaced with its superselected version, derived using \eqref{eq:superselection rule}. Comparison between classical and quantum states only makes sense for fixed resources devoted to the estimation; hence, we keep the 
average energy $E$ fixed. For classical states, {and writing $E$ as a dimensionless photon number,} this requires $E=\sum_{i=0}^d\left|\alpha_i\right|^2$,  
which can be used to recast the states in the form
\eq{
\prod_{i=0}^d\ket{\eu^{\iu\phi_i}\alpha_i}=\eu^{-E/2}\sum_{N=0}^\infty\frac{\left(\sum_{i=0}^d\eu^{\iu\phi_i}\alpha_i\had_i\right)^N}{N!}\ket{\text{vac}}.
} 
From this we can immediately identify the Fock layers as
\eq{
\sqrt{p_N}\ket{\psi_N}&=\eu^{-E/2}\frac{\left(\sum_{i=0}^d\eu^{\iu\phi_i}\alpha_i\had_i\right)^N}{N!}\ket{\text{vac}}\\
&=\frac{\eu^{-E/2}}{\sqrt{N!}}\sum_{k_0+\cdots+k_d=N}\sqrt{\binom{N}{\pmb{k}}}\eu^{\iu\pmb{k}\cdot\pmb{\phi}}\prod_{i=0}^d\alpha_i^{k_i}\ket{\pmb{k}},
}  using the multinomial coefficients $\binom{N}{\pmb{k}}=\binom{N}{k_0,\cdots,k_d}$, {and the vector notation $\ket{\pmb{k}}=\bigotimes_{j=1}^d\ket{k_j}_j$}. One can use the identity $\braket{\pmb{k}}{\pmb{k^\prime}} =\pmb{\updelta}_{k_0 k_0^\prime}\cdots\pmb{\updelta}_{k_d k_d^\prime}$ to verify that the probabilities are Poisson-distributed in terms of the total energy:
\eq{
p_N=
\frac{E^N \eu^{-E}}{N!}.
} The first-derivative terms are given by
\eq{
\braket{\psi_N}{\partial_j\psi_N}&=\frac{1}{p_N}
\frac{\eu^{-E}}{N!}\sum_{k_0+\cdots+k_d=N}\iu k_j\binom{N}{\pmb{k}}\prod_{i=0}^d\left|\alpha_i\right|^{2k_i}\\
&=
\iu|\alpha_j|^2 NE^{-1}}
and the second-derivative terms by
\eq{\braket{\partial_i \psi_N}{\partial_j \psi_N}&=\frac{1}{p_N}\frac{\eu^{-E}}{N!}
\sum_{\pmb{k}}k_i k_j\binom{N}{k}
\prod_{i=0}^d\left|\alpha_i\right|^{2k_i}\\
&=|\alpha_i|^2|\alpha_j|^2N(N-1)
E^{-2}}
 for  ($i\neq j$) and
\eq{\braket{\partial_j \psi_N}{\partial_j\psi_N}=|\alpha_j|^4 N(N-1)
E^{-2} +|\alpha_j|^2 N E^{-1}
} otherwise. The QFIM thus has components
\eq{
\pmb{H}_{i,j}&=4\sum_{N=0}^\infty p_N\left(\pmb{\updelta}_{i,j}\left|\alpha_i\right|^2 N E^{-1}-\left|\alpha_i\right|^2\left|\alpha_j\right|^2 NE^{-2}\right)\\
&=4\left(\pmb{\updelta}_{i,j}\left|\alpha_i\right|^2-\frac{\left|\alpha_i\right|^2\left|\alpha_j\right|^2}{E} \right).
\label{eq:QFIM coherent without reference}
}

The absence of a phase reference has two main consequences on the relative QFIM \eqref{eq:QFIM coherent without reference}, with respect to its counterpart with a phase reference \eqref{eq:QFIM coherent with reference}. The first observation is that each component of the QFIM is decreased, which is a stronger result than the averaging over a global phase not increasing the amount of information present in the states. \add{This need be taken into account by schemes showing quantum advantages with mode-separable states \cite{Knottetal2016,Gagatsosetal2016,Gessneretal2018}; the nonclassicality within a single mode cannot be harnessed for enhanced phase estimation in the absence of strong reference beams \cite{Geetal2018}.} More crucially, the rank of the QFIM  diminishes to $d$ from the value $d+1$ allowed by the presence of a reference. The number of \add{independent} parameters that can ultimately be estimated is thereby reduced by one when all resources are taken into consideration \add{because the state does not independently depend on all $d+1$ parameters}.

This reduction of the rank demands \rem{a judgment call}\add{careful consideration} when identifying how the available information should be used. A standard procedure consists of selecting one of the modes (viz. mode 0) as a reference, and comparing the other $d$ phases to $\phi_0$. The vector of parameters to be estimated is then given by   the relative phases  $\delta_{0,i}=\phi_i-\phi_0$.
\add{This can be achieved mathematically by acting on the entire state with the operator $\exp\left(-\iu \phi_0\hat{N}\right)$, which sends each phase $\phi_i\to\delta_{0,i}$ while leaving unchanged the final state in \eqref{eq:superselection rule}.} 
\add{
It can equivalently be achieved using the change-of-parametrization rule $\pmb{H}\to \pmb{J}_{1\to 0}^T\pmb{H}\pmb{J}_{1\to 0}$ with the $(d+1)\times (d+1)$ Jacobian that differentiates the original parameters with respect to the new:
\eq{
(\pmb{J}_{1\to 0})_{i,j}=\begin{cases}
\frac{\partial \phi_i}{\partial \delta_{0,j}} & j>0 \\
\frac{\partial \phi_i}{\partial \phi_0} & j=0
\end{cases};
} the relations $\phi_j=\delta_{0,j}+\phi_0$ and $\sum_{j=0}^d\pmb{H}_{i,j}=0$ ensure the new QFIM to have its first row and column vanish and the same final $d\times d$ block as the original QFIM.\footnote{
\add{We achieve the same QFIM with any global phase $\varphi$ such that $\phi_j=\delta_{0,j}+\varphi$. Then
$(\pmb{J}_{1\to 0})_{i,j}=\tfrac{\partial \phi_i}{\partial \delta_{0,j}}=\tfrac{\partial \delta_{0,i}}{\partial \delta_{0,j}}=\pmb{\delta}_{i,j}$
and 
$(\pmb{J}_{1\to 0})_{i,0}=\tfrac{\partial \phi_i}{\partial \varphi}=\tfrac{\partial \varphi}{\partial \varphi}=1$.
}
}
The block-diagonal $\pmb{H}$ allows us to retain all of the information present in the state by inspecting only the $d$ relative phase parameters \cite{Proctoretal2017}. 
Any other set of $d$ independent parameters that made $\pmb{H}$ block diagonal could have been chosen as a starting point because the Jacobians that transform between sets of $d$ independent parameters are invertible. Namely, all of the relative phases in Fig. \ref{fig:schemes} are spanned by the $d$ relative phases $\delta_{0,i}$ such that a change of \textit{parametrization} to another set of $d$ relative phases constitutes a change of \textit{basis}, up to some scale factor; the basis in which the QFIM is inverted has no effect on the optimal probe state nor the optimal measurement strategy. The cost matrix formalism can thus be used in this basis to ascertain the optimal probe states for \textit{any} set of relative-phase parameters.

}

All of the information about these $d$ relative phases is thus contained in the $d\times d$ submatrix of $\pmb{H}$ restricted to $i,j>0$. 
The inverse of the restricted $\pmb{H}$ can be computed using the Sherman-Morrison formula and has components
\eq{
\left(\pmb{H}^{-1}\right)_{i,j}=\frac{\pmb{\updelta}_{i,j}}{4\left|\alpha_i\right|^2}+\frac{1}{4\left|\alpha_0\right|^2},
\label{eq:inverse QFIM coh}
} 
with each diagonal term $\pmb{H}_{i,i}$ bounding from below the attainable uncertainty on the respective phase $\delta_{0,i}$. 

Selecting mode 0 as the reference does not
{yield the same result for the components of the QFIM corresponding to the $d$ phases $\delta_{0,i}$ that we saw earlier in Eq.~\eqref{eq:QFIM coherent with reference}. The correct form has smaller diagonal elements, accompanied by non-vanishing off-diagonal elements, which indicate the statistical correlations between the $d$ parameters.}
This is due to the fact that the expression \eqref{eq:QFIM coherent without reference} takes the finite energy in the reference mode into account explicitly; the ideal case is obtained only in the limit $\left|\alpha_0\right|^2\to\infty$.

To simultaneously optimize the estimation of all $d$ phases relative to a single phase reference one uses the cost matrix $\mathbf{R}_0=\id$ (\add{i.e., no change of parametrization; }see Fig. \ref{fig:schemes}) \add{to quantify a lower bound for the scalar quantum Cram\'er-Rao inequality \eqref{eq:scalar qCRB}}:
\eq{S_i=\tr\left(\mathbf{R}_i \pmb{H}^{-1}\right). \label{eq:general cost function}} 
The bound $S_0$ can be minimized using the Lagrange multiplier $E-\sum_{i=0}^d\left|\alpha_i\right|^2$ to enforce the constraint on the total energy. This yields
\eq{S_0=\frac{d}{4E}\left(\sqrt{d}+1\right)^2\label{eq:S0 classical}} when
\eq{\left|\alpha_i\right|^2=\frac{\left|\alpha_0\right|^2}{\sqrt{d}}=\frac{E}{d+\sqrt{d}}.\label{eq:optimal alpha original}}
We can compare this limit to what is attained performing $d$ sequential estimations \add{with the total energy of the reference mode taken into account; then, each estimation has} \rem{, each with} optimal energy $\left|\alpha_i\right|^2=\left|\alpha_0\right|^2=\tfrac{1}{2}\tfrac{E}{d}$ \add{such that each of the $d$ estimates uses a total of $\left|\alpha_0\right|^2+\left|\alpha_i\right|^2=\tfrac{E}{d}$ units of energy}. In this case, we find that that the analogue limit is $S_0=\tfrac{d^2}{E}$, thus larger by a factor of at most 4.
The imbalance between the optimal energy in the reference mode and the $d$ probe modes follows directly from its privileged position with respect to the estimation cost function.

\add{We can extend this to any weighting in the cost function: weigh each variance $\delta_{0,i}$ by $w_i$ such that $(\mathbf{R})_{i,j}=w_i \pmb{\delta}_{i,j}$. Then the optimal state has \eq{
E_i=\frac{E\sqrt{w_i}}{\sqrt{\sum_{i=1}^d w_i}+\sum_{i=1}^d \sqrt{w_i}};}
the optimal amount of energy in each component is exactly determined by the symmetries of the cost function.}

The choice of optimising the estimation of the parameters $\delta_{0,i}$ is not unique, nor necessarily the most convenient. In fact, one can imagine a symmetric situation in which the relevant quantities are the $d+1$ relative phases between each mode and the following one, parametrized by $\delta_{i,i+1}$, including $\delta_{d,0}$ (See Fig. \ref{fig:schemes}). Even though not all of the parameters are independent, \add{they all belong to the span of $\delta_{0,i}$, and }one can still equally weigh\rem{t} the cost of estimating each one. The variance of any relative phase can be determined from the covariance matrix of the original parametrization through
\eq
{
\Delta^2\left(\delta_{i,j}\right)=\Delta^2\left(\delta_{0,i}\right)+\Delta^2\left(\delta_{0,j}\right)-2\text{Cov}\left(\delta_{0,i},\delta_{0,j}\right),
}
where $\Delta^2\left(X\right)=\expct{X^2}-\expct{X}^2$
. Using $\Delta^2\left(\delta_{0,0}\right)=0$ and $\text{Cov}\left(\delta_{0,0},\delta_{0,d}\right)=\Delta^2\left(\delta_{0,d}\right)$, the minimum total uncertainty is bounded by
\begin{widetext}
\eq{
\Delta^2\left(\delta_{d,0}\right)+\sum_{i=0}^{d-1}\Delta^2\left(\delta_{i,i+1}\right)=2\left(\sum_{i=1}^{d}\Delta^2\left(\delta_{0,i}\right)-\sum_{i=1}^{d-1}\text{Cov}\left(\delta_{0,i},\delta_{0,i+1}\right)\right)\geq S_1=\tr\left(\mathbf{R}_1 \pmb{H}^{-1}\right)
\label{eq:ring cost variance sum}
}
\end{widetext}
for cost matrix \eq{\mathbf{R}_1=\begin{pmatrix}2&-1&0&\cdots&0\\
-1&2&-1&\cdots&0\\
\vdots&\ddots&\ddots&\ddots&\vdots\\
0&\cdots&-1&2&-1\\
0&\cdots&0&-1&2\end{pmatrix}.
} This cost matrix, and all others, can be found by the change-of-parametrization Jacobian \rem{$\pmb{J}_i$}\add{$\pmb{J}_{0\to i}$} through $\mathbf{R}_i=\rem{\pmb{J}_i^T\pmb{J}_i} \add{\pmb{J}_{0\to i}^T\pmb{J}_{0\to i}}$; here
\eq{
\rem{\pmb{J}_{i}^T}\add{\pmb{J}_{0\to 1}^T}=\left(
\begin{array}{cccccc}
 1 & -1 & 0 & 0 & \cdots & 0 \\
 0 & 1 & -1 & 0 & \cdots & 0 \\
 \vdots & \ddots & \ddots & \ddots & \ddots & \vdots \\
 0 & \cdots & 0 & 1 & -1 & 0 \\
 0 & \cdots & 0 & 0 & 1 & -1 \\
\end{array}
\right) \rem{.} 
} is found from taking derivatives of the new parameters with respect to the original ones:
\eq{
\left(\pmb{J}_{0\to 1}\right)_{i,j}=\frac{\partial \delta_{i,i+1} }{\partial \delta_{0,j}}=\frac{\partial \left(\delta_{0,i+1}-\delta_{0,i}\right) }{\partial \delta_{0,j}} .
}

The measure to be minimized for the ring cost function is \eq{
S_1=\frac{1}{2}\left(\frac{1}{\left|\alpha_0\right|^2}+\sum_{i=1}^d\frac{1}{\left|\alpha_i\right|^2}\right).
\label{eq:S1 classical}
} Using the same Lagrange multiplier as before, the optimal state now has \eq{
\left|\alpha_i\right|^2=\left|\alpha_0\right|^2=\frac{E}{d+1},
\label{eq:optimal alpha ring}
} corresponding to a lower-bounded total uncertainty of\footnote{Incorporating the weights $w_i$ on the variances $\delta_{i,i+1}$, the optimal state has $E_i=E\left(w_{i-1}+w_{i}\right)/2\sum_{i=0}^d w_i$ for all $i$ including $i=0$.} 
\eq{S_1=\frac{\left(d+1\right)^2}{2E}.} 

\textcolor{black}{In comparison, a sequential estimation scheme measures all $d+1$ parameters independently using energy $\tfrac{E}{d+1}$, with $(d+1)\times(d+1)$ QFIM equal to $\tfrac{E}{d+1}$ times the identity matrix. This can be recast as an estimate of the original $d$ independent parameters through the Jacobian transformation \eq{
\pmb{H}\to \rem{\pmb{J}_1^T\pmb{H}\pmb{J}_1}\add{\pmb{J}_{0\to 1}^T\pmb{H}\pmb{J}_{0\to 1}}=\frac{E}{d+1}\mathbf{R}_1 \rem{.}
} \add{because the QFIM transforms by differentiating the \textit{old} parameters with respect to the \textit{new} ones \cite{SidhuKok2020}, and the roles of \textit{old} and \textit{new} are reversed relative to when we derived the cost matrix}. The total error on all $d+1$ parameters for the sequential estimation scheme is easily calculated:
\eq{
S_1=\tr \left(\mathbf{R}_1\pmb{H}^{-1}\right)=\frac{d(d+1)}{E}.
} This is clearly superior to the alternative sequential estimation scheme in which only the original $d$ parameters are estimated; that case would have $S_1=\tr(\rem{R}\add{\mathbf{R}}_1 \tfrac{d}{E})=\tfrac{2d^2}{E}$.
}

\textcolor{black}{The simultaneous estimation strategy outperforms the sequential strategy by $\tfrac{2d}{d+1}$. The advantage approaches $2$ in the large-$d$ limit because each sequential estimation with energy $\tfrac{E}{d+1}$ can only send $\tfrac{E}{2(d+1)}$ through each mode, while our simultaneous estimation strategy always sends energy $\tfrac{E}{d+1}$ through each mode. Simultaneous estimation schemes are optimal due to their sharing of resources to minimize the variance in estimating each phase.}

One can finally consider the fully symmetric cost function that minimizes the sum of all pairwise relative phases \add{(again, spanned by the original $\delta_{0,i}$)} using
\eq{
\mathbf{R}_2=\begin{pmatrix}d&-1&-1&\cdots&-1\\
-1&d&-1&\cdots&-1\\
\vdots&\ddots&\ddots&\ddots&\vdots\\
-1&\cdots&-1&d&-1\\
-1&\cdots&-1&-1&d\end{pmatrix} \rem{.},
}
\add{where, as before, $\mathbf{R}_2=\pmb{J}_{0\to 2}^T\pmb{J}_{0\to 2}$, 
\eq{
\left(\pmb{J}_{0\to 2}\right)_{\mu,j}=\frac{\partial \delta_{\mu}}{\partial \delta_{0,j}}=\frac{\partial \left(\delta_{0,\mu_2}-\delta_{0,\mu_1}\right)}{\partial \delta_{0,j}},
} and $\mu$ indexes all of the pairs of relative phases as depicted in Fig. \ref{fig:schemes}.
}
It turns out that \eq{
S_2=\frac{d}{2}S_1,
} meaning that this measure is optimized by the same equal-energy state as the ring cost function. 
The optimal sequential estimation scheme, which measures $\binom{d+1}{2}$ relative phases each using total energy $E/\binom{d+1}{2}$\textcolor{black}{, also satisfies $S_2=\tfrac{d}{2}S_1$:
\eq{
S_2=d\frac{\binom{d+1}{2}}{E}.
}
We observe \textit{identical} behaviour in the ring and fully-connected parametrizations (Fig. \ref{fig:schemes}). This shows the generic result that symmetric estimation schemes are optimized by symmetric probe states, and that simultaneous estimation schemes can outperform sequential ones; we have seen this to be true for classical input states, and we will subsequently show the same phenomenon with quantum states.
}

\section{Optimal estimation with quantum states}

Nonclassical states are known to outperform their classical counterparts in phase estimation. This result holds in the single-parameter case, and carries over to the scenario of $d$ relative phases analogous to \eqref{eq:S0 classical}. For the latter case, the class of quantum states \eq{\ket{\psi}=\bigoplus_{i=0}^d\beta_i\ket{N}_i,
\label{eq:fixed N states}
} where the state $\ket{N}_i$ has $N$ photons in mode $i$ and zero photons elsewhere, has provided interesting insight in this problem~\cite{Humphreysetal2013,Pezzeetal2017}. This represents a generalization of $N00N$ states to our multidimensional problem, and, since the total photon number is fixed at the value $N$, it is left invariant under the superselection rule ~\eqref{eq:superselection rule}; the energy of the state is constrained to $E=N$ in units of photon number.

The QFIM for such states $\ket{\psi}$, which evolve to $\bigoplus_{i=0}^d\eu^{N\iu\phi_i}\beta_i\ket{N}_i$, has components
\eq{
\pmb{H}_{i,j}&=4N^2\left(\pmb{\updelta}_{i,j}\left|\beta_i\right|^2-\left|\beta_i\right|^2\left|\beta_j\right|^2\right)\\
&=4N\left(\pmb{\updelta}_{i,j}\left|\alpha_i\right|^2-\frac{\left|\alpha_i\right|^2\left|\alpha_j\right|^2}{N}\right),
\label{eq:QFIM NOONlike}
}
where we have defined the energy fraction for each mode to be $|\alpha_i|^2=N |\beta_i|^2$, in analogy with the classical case. This shows that the quantum expression for the QFIM \eqref{eq:QFIM NOONlike} differs from the classical one \eqref{eq:QFIM coherent without reference} only by the prefactor $N$; because of this form, all of the nuances from the classical treatment hold true -- except for the scaling advantage associated with the quantum resources. In particular, the QFIM again has rank $d$, making it necessary to select a strategy at the outset.

The first example on which we report, which has been discussed in \cite{Humphreysetal2013}, focuses on the $d$ phase differences $\delta_{0,i}$. The inverse of the QFIM is easily found to be
\eq{
\left(\pmb{H}^{-1}\right)_{i,j}=\frac{\pmb{\updelta}_{i,j}}{4N^2\left|\beta_i\right|^2}+\frac{1}{4N^2\left|\beta_0\right|^2}. \label{eq: S0 quantum}
} All of the optimization over $\left\{\beta_i\right\}$ carries through in exactly the same manner as in the classical case. In particular, the optimisation for the case of a single reference mode yields a total variance 
\eq{S_0=\frac{d}{4N^2}\left(\sqrt{d}+1\right)^2,} achieved for
$\left|\beta_i\right|^2=\left|\beta_0\right|^2/\sqrt{d}=\left(d+\sqrt{d}\right)^{-1}$ [c.f. \eqref{eq:S0 classical}], as discussed in \cite{Humphreysetal2013}. The scaling with the number of parameters $d$ is the same as for the classical simultaneous estimation strategy, while the scaling with energy is enhanced with these nonclassical states.

In comparison, the optimal sequential quantum estimation scheme measures the $d$ relative phases $\delta_{0,i}$ 
using a series of $d$ $N00N$ states, each with total photon number $N/d$. 
The precision on such a measurement is bounded by $\Delta^2\left(\delta_{0,i}\right)\geq 1/\left(N/d\right)^2$, thus yielding a  bound on the total precision \eq{\sum_{i=1}^d\Delta^2\left(\delta_{0,i}\right)\geq
\frac{d^3}{N^2}=\mathcal{O}\left(d\right)S_0.
} The simultaneous estimation scheme offers an enhancement over the sequential one by a factor that grows linearly with the number of phases being estimated. This growth with $d$, in contrast to the asymptotically-constant improvement of simultaneous versus sequential estimation with classical states, is due to the $E^{-2}$ scaling in the quantum case and the $E^{-1}$ scaling in the classical case. When the energy must be split into $d$ parts for a sequential estimation, the former suffers more than the latter\add{: sequential estimation increasing the total variance relative to that of simultaneous estimation by the order of $d^2$ in the quantum case, but only by the order of $d$ in the classical one}. \add{The extra multipartite correlations make}\rem{This makes} simultaneous estimation much more appealing in the quantum case.

Moving beyond the picture of a privileged phase-reference mode, we again consider the ring cost function \eqref{eq:ring cost variance sum}. We find the optimal states to be
GHZ-type states defined by $\left|\beta_i\right|^2=\left|\beta_0\right|^2=\left(d+1\right)^{-1}$, with [c.f. \eqref{eq:S1 classical} and \eqref{eq:optimal alpha ring}] \eq{S_1=\frac{\left(d+1\right)^2}{2N^2}.}\add{ Because the QFIM is degenerate with respect to the phase differences being calculated, the same result is obtained regardless the set of $d$ independent relative phases with which we begin.}
This can again be compared to the optimal sequential estimation scheme using $d+1$ $N00N$ states with $N/(d+1)$ photons each, achieving a lower-bounded sum of variances \eq{
\sum_{i=0}^d \Delta^2(\delta_{i,i+1})\geq d\left(\frac{d+1}{N}\right)^2=\mathcal{O}\left(d\right)S_1.
} The increased advantage with $d$ has the same origin in the different energy scaling as the previous case. \textcolor{black}{A sequential estimation of only $d$ independent parameters, on the other hand, gives $\tr\left[\mathbf{R}_1\left(\tfrac{d}{N}\right)^2\right]=2d\left(\tfrac{d}{N}\right)^2$, which outperforms the sequential ring estimation scheme for $d<3$.}

Lastly we turn our attention to the fully-symmetric cost of estimating all $\binom{d+1}{2}$ phase differences. The optimization for the classical states carries through to again be optimized by the GHZ-type states, with 
\eq{S_2=\frac{d}{2}S_1=\frac{d\left(d+1\right)^2}{4N^2}.
} A sequential estimation strategy this time requires $\binom{d+1}{2}$ $N00N$ states with $N/\binom{d+1}{2}$ photons each, resulting in a total sum of variances \textcolor{black}{\eq{
\sum_{i<j=0}^d\Delta^2(\delta_{i,j})\geq 
d\left(\frac{\binom{d+1}{2}}{N}\right)^2=\mathcal{O}\left(d^2\right)S_2.
} This sequential estimation scheme performs poorly. Better is to only perform a sequential estimation of $d$ phases relative to a single common reference, and to infer the values of the other parameters; this procedure has variance bound $\tr\left[\mathbf{R}_2 \left(\frac{d}{N}\right)^2\right]=\tfrac{d^4}{N^2}$. The ring estimation procedure, with variance bound  $\tr(\mathbf{R}_2\mathbf{R}_1^{-1})\tfrac{(d+1)^2}{N^2}=\tfrac{1}{2}\binom{d+2}{3}\tfrac{(d+1)^2}{N^2}$, outperforms both others for $d=3,4$ by less than $5\%$; otherwise, it is better to avoid splitting the energy $N$ into too many parts due to the $\mathcal{O}(N^2)$ scaling of the variances.}

We can make some overall comments about uncertainty scalings\textcolor{black}{, with results summarized in Table \ref{tab:variances table}}. The optimal simultaneous quantum estimation strategy goes as
\eq{\sum_{i,j}\Delta^2(\delta_{i,j})\geq \mathcal{O}\left(ndN^{-2}\right),} where $n$ is the total number of possibly-dependent parameters being estimated (i.e., the total number of terms in the sum over $i,j$). This is because the variance goes as $\sim\mathcal{O}(E^{-2})=\mathcal{O}(N^{-2})$, the probabilities $|\beta_i|^2$ are equally split among $\mathcal{O}(d)$ modes, and there are $\mathcal{O}(n)$ total covariance terms to consider. \textcolor{black}{In contrast, for a sequential quantum scaling the energy is optimally split into $\mathcal{O}(d)$ parts with probabilities split over only $\mathcal{O}(1)$ modes, and there are $\mathcal{O}(n)$ variances to sum, leading to an overall scaling \eq{
\sum_{i,j}^d\Delta^2(\delta_{i,j})\geq 
\mathcal{O}\left[n\left(\frac{d}{N}\right)^2\right]=\mathcal{O}\left(nd^2 N^{-2}\right).}} 

For a classical simultaneous estimation there are again $n$ parameters to be estimated, this time with variances scaling as $E^{-1}$, and the total energy is again split among $d$ modes:
\eq{\sum_{i,j}\Delta^2(\delta_{i,j})\geq \mathcal{O}\left(n d E^{-1}\right).} The sequential classical scheme divides the energy into $n$ parts, but the reparametrized information only counts these $d$ times, leading to the same scaling. Simultaneous estimation outperforms sequential estimation in the classical regime by $\mathcal{O}(1)$ due to advantages in resource allocation among modes.
These scaling arguments explain the asymptotic improvements of the simultaneous estimation schemes for different numbers of parameters being estimated\textcolor{black}{. Quantum schemes outperform their classical counterparts by a factor of $N$; this heightened sensitivity to splitting $N$ among more measurements is responsible for the dramatic $\mathcal{O}(d)$ improvements promised by simultaneous versus sequential quantum estimation.}

\begin{table*}[]
\centering
\caption{Minimum total variances $\Delta^2_\text{tot}$ for each estimation scheme depicted in Fig. \ref{fig:schemes}. For all configurations, simultaneous quantum estimation schemes outperform sequential and classical estimation ones. Classical sequential estimation schemes benefit from estimating each parameter directly, even those are are not independent; quantum simultaneous estimation schemes are better-served by measuring fewer parameters. \add{Changing the relative significance of each parameter in the estimation procedure changes the optimal strategy accordingly.}}
\label{tab:variances table}
\resizebox{\textwidth}{!}{%
\begin{tabular}{@{}
>{\columncolor[HTML]{FFFFFF}}c 
>{\columncolor[HTML]{FFFFFF}}c 
>{\columncolor[HTML]{EFEFEF}}c 
>{\columncolor[HTML]{FFFFFF}}c 
>{\columncolor[HTML]{EFEFEF}}c 
>{\columncolor[HTML]{FFFFFF}}c 
>{\columncolor[HTML]{EFEFEF}}c 
>{\columncolor[HTML]{FFFFFF}}c @{}}
\toprule
\multicolumn{2}{c}{\cellcolor[HTML]{FFFFFF}} &
  \multicolumn{2}{c}{\cellcolor[HTML]{FFFFFF}Common reference ($\mathbf{R}_0$)} &
  \multicolumn{2}{c}{\cellcolor[HTML]{FFFFFF}Neighbouring references ($\mathbf{R}_1$)} &
  \multicolumn{2}{c}{\cellcolor[HTML]{FFFFFF}All references ($\mathbf{R}_2$)} \\ \cmidrule(l){3-8} 
\multicolumn{2}{c}{\multirow{-2}{*}{\cellcolor[HTML]{FFFFFF}}} &
  \cellcolor[HTML]{FFFFFF}$\Delta^2_\text{tot}$ &
  Strategy &
  \cellcolor[HTML]{FFFFFF}$\Delta^2_\text{tot}$ &
  Strategy &
  \cellcolor[HTML]{FFFFFF}$\Delta^2_\text{tot}$ &
  Strategy \\ \midrule
\begin{tabular}[c]{@{}c@{}}\textcolor{white}{z}\\ \textcolor{white}{z}\\ Classical\end{tabular} &
  \begin{tabular}[c]{@{}c@{}}\textcolor{white}{z}\\ Sequential\\ \textcolor{white}{z}\end{tabular} &
  $\frac{d^2}{E}$ &
  $d$ estimates &
  $\frac{d(d+1)}{E}$ &
  $d+1$ estimates &
  $\frac{d^2(d+1)}{2E}$ &
  $\binom{d+1}{2}$ estimates \\ \cmidrule(l){2-8} 
 &
  \begin{tabular}[c]{@{}c@{}}\textcolor{white}{z}\\ Simultaneous\\ \textcolor{white}{z}\end{tabular} &
  $\frac{d(\sqrt{d}+1)^2}{4E}$ &
  privileged mode &
  $\frac{(d+1)^2}{2E}$ &
  mode symmetry &
  $\frac{d(d+1)^2}{4E}$ &
  mode symmetry \\ \midrule
\begin{tabular}[c]{@{}c@{}}\textcolor{white}{z}\\ \textcolor{white}{z}\\ Quantum\end{tabular} &
  \begin{tabular}[c]{@{}c@{}}\textcolor{white}{z}\\ Sequential\\ \textcolor{white}{z}\end{tabular} &
  $\frac{d^3}{N^2}$ &
  $d$ estimates &
  \Bigg\{
  \begin{tabular}[c]{@{}c@{}}$\frac{2d^3}{N^2}$,  $d\leq 2$\\ $\frac{d(d+1)^2}{N^2}$, $d>2$\end{tabular} &
  \begin{tabular}[c]{@{}c@{}}$d$ estimates\\ $d+1$ estimates\end{tabular} &
  \Bigg\{
  \begin{tabular}[c]{@{}c@{}}$\binom{d+2}{3}\frac{(d+1)^2}{2N^2}$, $d=3,4$\\ $\frac{d^4}{N^2}$, $d\neq3,4$\end{tabular} &
  \begin{tabular}[c]{@{}c@{}}$d+1$ estimates\\ $d$ estimates\end{tabular} \\ \cmidrule(l){2-8} 
 &
  \begin{tabular}[c]{@{}c@{}}\textcolor{white}{z}\\ Simultaneous\\ \textcolor{white}{z}\end{tabular} &
  $\frac{d(\sqrt{d}+1)^2}{4N^2}$ &
  privileged mode &
  $\frac{(d+1)^2}{2N^2}$ &
  mode symmetry &
  $\frac{d(d+1)^2}{4N^2}$ &
  mode symmetry \\ \bottomrule
\end{tabular}%
}
\end{table*}

\section{Optimal measurement scheme for simultaneous phase estimation}

The matrix quantum Cram\'er-Rao bound suffers from the limitation that the bound may be unattainable even in principle. Multiple phase estimation is a fortunate counter-example that does not suffer from this drawback; it is possible to simultaneously estimate all $d$ independent parameters. This is ultimately linked to the fact the generators corresponding to the $d$ independent phase shift operations commute. For example, the commutativity of the set \eq{\hat{h}_{\delta_{0,i}}=\hat{n}_i\rem{-\hat{n}_0},\quad i\in(1,d)} implies that the $d$ phase differences $\delta_{0,i}$ can be simultaneously estimated at the ultimate limit. 

We present example schemes that can be experimentally implemented to saturate the Cram\'er-Rao inequality \eqref{eq:qCRB matrix}. Notice that, while the QFIM is only $d$-dimensional, the quantum Cram\'er-Rao bound can be saturated for any cost function \eqref{eq:general cost function}, even one that takes into account more than $d$ interdependent parameters. We focus our discussion on quantum states with fixed $N$, as these are the only states that remain pure in the absence of a phase reference.

Per Refs. \cite{Humphreysetal2013,Pezzeetal2017}, the quantum Cram\'er-Rao bound can be saturated by a projection-valued measure using of a set of $d+1$ orthogonal projectors. These depend explicitly on the phases $\pmb{\phi}$, which \rem{this} is perfectly legitimate, since the Cram\'er-Rao bound holds true for local estimation and in the asymptotic limit. The first projector is chosen to correspond to the evolved state $\ket{\psi\left(\pmb{\phi}\right)}$. Since this state and its $d$ derivatives \eq{
\frac{\partial}{\partial \delta_{0,i}}\ket{\psi\left(\pmb{\phi}\right)}=\iu\hat{h}_{\delta_{0,i}}\ket{\psi\left(\pmb{\phi}\right)}
} are linearly independent, the remaining $d$ projectors can be formed using a Gram-Schmidt orthogonalization procedure, provided they satisfy some additional conditions detailed in Ref.~\cite{Pezzeetal2017}. 
\add{Because all of the generators $\hat{h}_{\delta_{i,j}}$ of the relative phases can be created from linear combinations of the original $d$ generators $\hat{h}_{\delta_{0,i}}$, the orthogonalization procedure does not actually depend on which $d$ independent relative phases one asserts to estimate.}

The first scenario in Fig. \ref{fig:schemes}a corresponds to $d$ phase shifts $\delta_{0,i}$ being estimated with respect to a common reference. The state identified by the condition \eqref{eq: S0 quantum} represents the optimal choice. In the limit of small phase shifts $\pmb{\phi} \approx 0$, which can be obtained by means of adaptive schemes, the optimal measurement includes $d+1$ projectors
\eq{
\ket{\mathbf{u}^{(j)}}=\bigoplus_{i=0}^d u_i^{(j)}\ket{N}_i,\quad j\in(1,{\color{black}d+1}).
}
{\color{black}To obtain the coefficients $u_i^{(j)}$ ~\cite{Humphreysetal2013}, where the first vector must be the projector over the unperturbed input state, one can define the following set of linearly independent vectors:
\begin{equation}
\label{eq:Humphreys_meas}
\begin{cases}
\mathbf{v}^{(1)} \propto (d^{1/4}, 1, \ldots, 1), \\
\mathbf{v}^{(2)} \propto (d^{1/4}, 1, \ldots, 1, -1), \\
\mathbf{v}^{(3)} \propto (d^{1/4}, 1, \ldots, 1, -1, 0), \\
\mathbf{v}^{(4)} \propto (d^{1/4}, 1, \ldots, 1, -1, 0, 0), \\
\ldots\\
\mathbf{v}^{(d+1)} \propto (d^{1/4}, -1, 0, \ldots, 0).
\end{cases}
\end{equation}
Note that first vector $\mathbf{v}^{(1)}$ is, modulo a normalization constant, the projector onto the unperturbed probe state. Then, by applying Gram-Schmidt orthogonalization to the set $\left\{\mathbf{v}^{(j)}\right\}$, one finds a novel set of vectors $\left\{\mathbf {u}^{(j)}\right\}$ that provides the coefficients for states $\ket{\mathbf{u}^{(j)}}$. Such a choice leads to the same set of projectors reported in \cite{Humphreysetal2013} for the specific case $d=3$, and provides a general recipe for larger $d$}. {\color{black}Furthermore}, such a choice is guaranteed to satisfy the conditions of \cite{Pezzeetal2017}, since all coefficients are real-valued. {\color{black}Finally}, this specific set maximizes the number of zero-valued coefficients, potentially simplifying an experimental apparatus performing such measurement.

An analogous set can be obtained for the GHZ-type states that are optimal for the ring and fully-connected cost functions (Fig. \ref{fig:schemes}b-c). In these cases, in the limit of small phase shifts $\pmb{\phi} \approx 0$, {\color{black}the set of projectors $\ket{\mathbf{u}^{(j)}}$ is obtained by applying Gram-Schmidt orthogonalization to the following vectors:
\begin{equation}
\label{eq:GHZ_meas}
\begin{cases}
\mathbf{v}^{(1)} = (1, 1, \ldots, 1, 1), \\
\mathbf{v}^{(2)} = (1, 1, \ldots, 1, -1), \\
\mathbf{v}^{(3)} = (1, 1, \ldots, 1, -1, 0), \\
\mathbf{v}^{(4)} = (1, 1, \ldots, 1, -1, 0, 0), \\
\ldots\\
\mathbf{v}^{(d+1)} = (1, -1, 0, \ldots, 0),
\end{cases}
\end{equation}
which leads to a novel set of orthonormal vectors $\left\{\mathbf {u}^{(j)}\right\}$ defining the measurement. As for the previous scenario with $d$ phase shifts measured with respect the common mode $0$, the first vector $\left\{\mathbf{v}^{(1)}\right\}$, and thus $\left\{\mathbf{v}^{(1)}\right\}$ after Gram-Schmidt orthogonalization, is the projector onto the unperturbed probe state.}
As before, such coefficients define the states {\color{black}$
\ket{\mathbf{u}^{(j)}}=\bigoplus_{i=0}^d u_i^{(j)}\ket{N}_i$}, {\color{black}and lead to the same results as Ref. \cite{Humphreysetal2013}.}

Note that, in general, the set of projectors for a given probe state is not unique. As an explicit example, let us again consider the GHZ-type states for the specific case $d=3$. In addition to the optimal measurement constructed from \eqref{eq:GHZ_meas}, a different set of projectors satisfying the conditions of~\cite{Pezzeetal2017} can be obtained as:
\begin{equation}
\begin{cases}
\mathbf{u^{\prime}}^{(1)} = \frac{1}{2} (1, 1, 1, 1),\\
\mathbf{u^{\prime}}^{(2)} = \frac{1}{2} (1,-1, 1, -1), \\
\mathbf{u^{\prime}}^{(3)} = \frac{1}{2} (1, 1, -1, -1), \\
\mathbf{u^{\prime}}^{(4)} = \frac{1}{2} (1, -1, -1, 1).
\end{cases}
\end{equation}
One may thus use whichever set of optimal projectors best suits their experimental constraints.

For a given probe state, the optimal measurement procedure does not depend on the cost function. In order to show this, we consider a specific measurement saturating the quantum Cram\'er-Rao $\pmb{C} \geq \pmb{H}^{-1}$. When a generic cost matrix $\mathbf{R}_{i}$ is introduced, the corresponding bound on the cost function is calculated by evaluating both the covariance matrix $\pmb{C}$ and the QFIM $\pmb{H}$ in conjunction with the same $\mathbf{R}_{i}$, namely \add{\eqref{eq:scalar qCRB}} $\mathrm{Tr}(\mathbf{R}_{i} \pmb C) \geq \mathrm{Tr}(\mathbf{R}_{i} \pmb{H}^{-1})$. The saturation of the quantum Cram\'er-Rao bound then implies saturation of the bound subject to the specific cost function. In turn, this means that, given a fixed state and a measurement saturating the quantum Cram\'er-Rao bound, this measurement is capable of optimally extracting information on any set of parameters  specified by an arbitrary  $\mathbf{R}_{i}$. 

The optimal measurement scheme is easy to depict for $N=1$. In that case, the optimal probe state can be generated by passing a single photon through a series of beam splitters and phase elements; then, after the application of the phases $\pmb{\phi}$, a projection onto the probe state can be achieved by running the circuit in reverse and projecting onto the presence of a single photon in the original mode. The local asymptotic nature of this estimation scheme is exemplified by the reverse circuit requiring a good estimate $\tilde{\pmb{\phi}}$ of the phases. The remaining $d$ projectors, corresponding to detecting the photon in any of the $d$ remaining modes, will give zero probability in the limit $\tilde{\pmb{\phi}}\to \pmb{\phi}$. For $N>1$ photons, not all states and measurements can be deterministically accessed by linear-optical passive networks, and thus in general may require one to exploit additional ingredients such as post-selection~\cite{Pezzeetal2017}.

\section{Conclusions}
In metrology, the concept of resources is central to quantifying and comparing the aptness of different strategies. While phase measurements are often considered as illustrative and technologically relevant examples, care should be taken in stating what resources are actually employed, due to the rich conceptual intricacy; the inability of defining absolute phases is a notable case in point.

In this article we have focused on a comprehensive theory of how the unavailability of a phase reference limits multiphase estimation. In the absence of such an external reference, one has to optimize their measurements for a particular choice of relative phases. This choice then dictates the optimal probe state for a simultaneous estimation of all parameters.

We have first derived the general limits that can be attained by classical resources, and then we have introduced significantly improved quantum strategies based on quantum states of light with fixed numbers of photons. Within this class, states of the form \eqref{eq:fixed N states} undergo large phase shifts $N\phi_i$, leading to Heisenberg scaling of the Fisher information with average energy. Moreover, simultaneously estimating all of the parameters in symmetric estimation schemes is dramatically more precise than sequentially estimating each relative phase. The usefulness of this phenomenon will hopefully extend to a wide variety of quantum-enhanced multiparameter estimation problems.

Absent a privileged reference arm, there are multiple arrangements in which the phases can be self-referenced. We have explored three cases: selecting one arm as a reference; referring each arm to its neighbors; and considering all possible pairs of modes. Since the Cram\'er-Rao bound can be saturated in all of these cases, there is no preference for one over the others. It is important to remark how the symmetry, or lack thereof, of the different parameters is mirrored by the optimal state\add{, even when each parameter has arbitrary significance}. We expect that similar considerations of the role of symmetry can find applications in more general settings of multiparameter estimation.

\begin{acknowledgments}
AZG acknowledges funding from an NSERC Discovery
Award Fund, an NSERC
Alexander Graham Bell Scholarship, the Walter C.
Sumner Foundation, the Lachlan Gilchrist Fellowship Fund, a Michael Smith Foreign Study Supplement, and Mitacs Globalink. NS and FS acknowledge funding from MIUR via PRIN ``QUSHIP - Taming complexity with quantum strategies: a hybrid integrated photonics approach'' and by the Amaldi Research Center funded by the Ministero
dell'Istruzione dell'Universit\`a e della Ricerca (Ministry of
Education, University and Research) program ``Dipartimento
di Eccellenza'' (CUP:B81I18001170001).  AMS is a Fellow of CIFAR, and further acknowledges support from an NSERC Discovery Grant and from a visiting professor fellowship of La Sapienza.
\end{acknowledgments}

%

\end{document}